\documentclass[prd,twocolumn,eqsecnum,showpacs]{revtex4}

\newcommand{\C}{\mathcal{C}}

\usepackage[dvips]{graphicx}
\usepackage{color}
\usepackage{amsmath}
\usepackage{amssymb}

\begin{document}
%
%
%
%
\preprint{LA-UR 05-0556}
\title
   {Renormalized broken-symmetry Schwinger-Dyson equations \\
    and the 2PI--1/N expansion for the O(N)  model}
\author{Fred~Cooper}
\email{fcooper@nsf.gov}
\affiliation{National Science Foundation,
   Division of Physics,
   Arlington, VA 22230}
\affiliation{Santa Fe Institute,
   Santa Fe, NM 87501}
\affiliation{Theoretical Division,
   Los Alamos National Laboratory,
   Los Alamos, NM 87545}
\author{John~F.~Dawson}
\email{john.dawson@unh.edu}
\affiliation{
   Department of Physics,
   University of New Hampshire,
   Durham, NH 03824}
\author{Bogdan~Mihaila}
\email{bmihaila@lanl.gov}
\affiliation{Theoretical Division,
   Los Alamos National Laboratory,
   Los Alamos, NM 87545}
\date{\today}
\begin{abstract}
We derive the renormalized Schwinger-Dyson equations for the one-
and two-point functions in the auxiliary field formulation of
Coleman, Jackiw, and Politzer~\cite{coleman} for $\lambda \, \phi^4$
field theory, to order 1/N, in the 2PI--1/N expansion.  We show that
the renormalization of the broken-symmetry theory depends only on
the counter terms of the symmetric theory with $\phi = 0$, as
discussed in our previous paper~\cite{renorm1}. We find that the
2PI--1/N expansion violates the Goldstone theorem at order 1/N. In
using the O(4) model as a low energy effective field theory of pions
to study the time evolution of disoriented chiral condensates one
has to {\em{explicitly}} break the O(4) symmetry to give the
physical pions a nonzero mass. In this effective theory, we expect
that the additional small contribution to the pion mass due to the
violation of the Goldstone theorem in the 2PI-1/N equations to be
unimportant for an adequate description of the phenomenology.
\end{abstract}
%
\maketitle

%
%
\section{Introduction}

Lately there has been interest in using two-particle irreducible
(2PI)--1/N methods to investigate various aspects of quantum field
theory~\cite{Aarts_66,2PI_1N}.  In a previous work~\cite{renorm1} we
showed how to renormalize the Schwinger-Dyson (SD) equations for the
symmetric phase of the O(N) model in the auxiliary field formalism,
to order 1/N. This was done by first using the multiplicative
renormalization approach~\cite{BCG} to find the exact renormalized
SD equations, and then to realize that to leading order in 1/N one
only needs to set the renormalized vertex function for $\phi \phi
\chi$ to one ($\Gamma_R = 1$), in order to consistently truncate the
infinite hierarchy of renormalized Green's functions. Here, $\chi$
is the auxiliary field related to $\phi^2$.

In order to carry out dynamical simulations with non-zero values of
$\langle \phi \rangle$, which occurs, for example, when chiral
condensates are produced~\cite{DCC}, it is important to extend that
result to the case of broken symmetry, $\phi \neq 0$. In this paper
we show that by extending the multiplicative renormalization scheme
used for the symmetric phase, we can  obtain finite renormalized S-D
equations for the broken symmetry phase. Since the 2PI--1/N approach
is a resummation of the ordinary 1/N expansion, it is important to
ask to what extent the Ward-Takahashi identities preserved in the
original perturbative 1/N approach~\cite{BCG} are preserved in the
2PI--1/N approach. One of these identities leads to the Goldstone
theorem~\cite{Goldstone}. Goldstone's theorem states that if
continuous symmetry is broken, and there is a residual symmetry,
there should be massless particles in the theory corresponding to
the number of symmetries left unbroken.

As has been previously pointed out for the 2PI--1/N, the Goldstone
theorem is formally satisfied if one determines the masses from the
inverse propagators derived from the one-particle irreducible (1PI)
generating functional for the $\phi$
fields~\cite{hees_02a,Aarts_66}. However one expects (and we find)
that the inverse propagators obtained directly from the 2PI--1/N
generating functional  for the would be massless particles do {\em
not} vanish as $p^2$ as $p^2 \rightarrow 0$ in violation of
Goldstone's theorem. What we explicitly find is that the condition
for the spontaneous symmetry breakdown found from the renormalized
equation for the expectation value of $\langle \phi_i \rangle$ leads
to a mass for the would be Goldstone bosons at order 1/N.  The
evolution equations obtained from the 2PI--1/N effective action are
energy preserving.

If we modify {by hand} these equations to enforce the Goldstone
theorem, we would then violate energy conservation at order 1/N.
This violation of the Goldstone theorem can be more satisfactorily
remedied by constructing an improved effective action functional as
discussed in Van Hees and Knoll~\cite{hees_02a}. However, this then
leads to a much more complicated set of equations which includes (in
addition to solving for the one and two point function equations)
solving simultaneously the Bethe-Salpeter equations for the vertex
function. Given present computational power, this would be not
feasible for $3+1$ dimensional calculations at this time. The
improved effective action is not entirely satisfactory in that the
propagators on internal lines still do not obey the Goldstone
theorem.

One can hope, however, from a phenomenological point of view, that
the violation of the Goldstone theorem by this approximation is not
very serious. In making a realistic phenomenological model of pions
using the O(4) model one has to \emph{explicitly} break the O(4)
symmetry if we want the pion to have the correct physical mass. This
is done  by introducing an external source coupling to the field
with non-vanishing expectation value (i.e. the $\sigma$ particle).
One then determines the magnitude of this external source by  using
the partially conserved axial current equation (PCAC). This was
discussed in a previous paper on disoriented chiral
condensates~\cite{DCC}.  As long as the mass generated by the
breakdown of the Goldstone theorem is small compared to the mass
generated by the {\em explicit} violation of the symmetry then this
breakdown should not be important for phenomenological applications.
Our renormalization procedure will lean heavily on our previous
result for obtaining renormalized S-D equations in the symmetric
vacuum~\cite{renorm1}.

Before continuing with our approach we will discuss some previous
approaches to Goldstone theory problems. Firstly, in a direct 1/N
expansion to order 1/N$^2$, Binoth \emph{et al}~\cite{Binoth} have
performed all renormalizations. They found no inconsistencies with
the Goldstone theorem, and the residual O(N-1) symmetry is
preserved. This is a very comforting result, but as we have
discussed previously~\cite{secularity}, a direct 1/N expansion leads
to secularity problems in the dynamics which is our main interest
here. In an important paper, Arrizabalaga  \emph{et al}~\cite{jan}
realized that one can avoid the problems with the Goldstone theorem
discussed here, by breaking the symmetry and then taking the limit
of zero symmetry-breaking. This is valid if we one is interested in
O(N)-invariant initial conditions, but also having Goldstone
particles. We will discuss this approach later. Finally, Ivanov
\emph{et al}~\cite{knoll05} have discussed how to preserve the
Goldstone theorem in the simpler Hartree approximation, by adding
terms to the 2PI generating functional which vanish when the
symmetry is restored, but which explicitly enforce the Goldstone
condition. This is a promising approach, which needs to be explored
further.

%
%
\section{The O(N) scalar field theory}

In the auxiliary field formulation of Coleman \emph{et
al}~\cite{coleman}, the O(N) model can be described by an action
written in terms of the auxiliary field~$\chi$
\begin{align}
   S [\phi_i, \chi]
   \ = \
      \int \mathrm{d}^d x \
   \biggl \{ &
   - \frac{1}{2} \,
   \phi_i(x) \,
   \bigl [ \Box + \chi(x) \bigr ] \, \phi_i(x)
   \notag \\ &
   +
   \frac{\chi^2(x)}{2 g}
   +
   \frac{\mu^2}{g} \chi(x) \,
   \biggr \} \>.  \label{action}
\end{align}
Here and in what follows we let $g=\lambda/N$. To treat the $N+1$
fields on equal footing we introduce the notation
\begin{equation}\label{gold.e:phijextended}
\begin{split}
   \phi_a(x)
   \ = \
   &   [ \, \chi(x), \phi_1(x), \phi_2(x), \ldots , \phi_N(x) \, ] \>,
   \\
   j_a(x)
   \ = \
   &   [ \, j_0(x), j_1(x), j_2(x), \ldots , j_N(x) \, ]  \>,
\end{split}
\end{equation}
with $a=0,i=1,\ldots,N$.  Using this notation, the complete action
for the O(N) model is given by:
\begin{multline}\label{gold.e:fullaction}
   S[\phi;j]
   \ = \
   -
   \frac{1}{2}
      \int \mathrm{d}^d x \!\! \int \mathrm{d}^d x' \,
   \phi_a(x) \, \Delta_{ab}^{-1}(x,x') \, \phi_b(x')
   \\
   +
   \int \mathrm{d}^d x \,
   \Bigl \{ \,
      -
      \frac{1}{6} \, \gamma_{abc} \,
      \phi_a(x) \, \phi_b(x) \, \phi_c(x)
      +
      \phi_a(x) \, j_a(x) \,
   \Bigr \} \>,
\end{multline}
where $\Delta_{ab}^{-1}(x,x') = \Delta_{ab}^{-1}(x) \, \delta(x,x')$
with
\begin{equation}\label{e:gold.Deltainvdef}
   \Delta_{ab}^{-1}(x)
   \ = \
      \begin{pmatrix}
      -1/g & 0                  \\
       0   & \Box \, \delta_{ij}
   \end{pmatrix} \>,
\end{equation}
and where $\gamma_{abc} = \delta_{a0} \delta_{ij} + \text{cyclic
permutations}$.  Here we have put $j_0(x) = J(x) + \mu^2 / g$.  The
coupling constant $g = \lambda/N$ is of order 1/N. For the dynamics
the integrals and delta functions $\delta_{\C}(x,x')$ are defined on
the closed time path (CTP) contour, which incorporates the initial
value boundary condition~\cite{CTP}.

The generating functional $Z[j]$ and connected Green's function
generator $W[j]$ are defined by a path integral:
\begin{equation}\label{gold.e:ZWdef}
   Z[j]
   \ = \
      e^{\mathrm{i} \, W[j] }
   \ = \
      \prod_{a=0}^{N} \int \mathrm{d} \phi_a \,
   e^{\mathrm{i} \, S[\phi;j]} \>.
\end{equation}
We define one-point functions by:
\begin{equation}\label{gold.e:onepointGreendef}
   \phi_a(x)
   \ = \
      \frac{\delta W[j]}{\delta j_a(x)} \>,
\end{equation}
which satisfy the equations:
\begin{multline}\label{gold.e:onepointGreeneq}
   \Delta_{ab}^{-1}(x) \, \phi_b(x)
   \\
   +
   \frac{1}{2} \, \gamma_{abc} \,
   \bigl \{ \,
      \phi_b(x) \, \phi_c(x)
      +
      G_{bc}(x,x)/\mathrm{i} \,
   \bigl \}
   \ = \
      j_a(x) \>,
\end{multline}
where $G_{ab}(x,x')$ is the two-point Green's function, defined by:
\begin{align}\label{gold.e:twopointGreendef}
   G_{ab}(x,x')
   \ = \
   &
      \frac{\delta \phi_a(x)}{\delta j_b(x')}
   \ = \
      \frac{\delta^2 W[j]}{\delta j_a(x)\,\delta j_b(x')}
   \\
   \ = \
   &
      \begin{pmatrix}
      D(x,x')     & K_j(x,x') \\
      \bar{K}_i(x,x') & G_{i j}(x,x')
   \end{pmatrix}  \>,
   \label{e:GGdef}
   \notag
\end{align}
We also define the generating functional $\Gamma[\phi]$ of 1PI
vertices by a Legendre transformation:
\begin{equation}
   \Gamma[\phi]
   \ = \
      W[j]
   -
      \int \mathrm{d}^d x \ \phi_a(x) j_a(x) \>,
\end{equation}
and one-point vertex functions by:
\begin{equation}
   \Gamma_a^{(1)}(x)
   \ = \
      - \frac{\delta \Gamma[\phi]}{\delta \phi_a(x)}
   \ = \
      j_a(x) \>,
\end{equation}
so that from \eqref{gold.e:onepointGreeneq}, we have:
\begin{multline}\label{gold.e:onepointvertexeq}
   \Gamma_a^{(1)}(x)
   \ = \
      \Delta_{ab}^{-1}(x) \, \phi_b(x)
   \\
   +
   \frac{1}{2} \, \gamma_{abc} \,
   \bigl \{ \,
      \phi_b(x) \, \phi_c(x)
      +
      G_{bc}(x,x)/\mathrm{i} \,
   \bigl \} \>,
\end{multline}
which gives the familiar equations of motion
\begin{equation}\label{eq_motion}
\begin{split}
   &
   \bigl [ \Box + \chi(x) \bigr ] \, \phi_i(x)
   +
   K_i(x,x) / i
   \ = \
   j_i(x) \>,
    \\
    &
   \chi(x)
   \ = \
   - \mu^2-g j_0(x)+
   \frac{g}{2}
      \sum_i
      \bigl [ \,
         \phi_i^2(x)
         +
         G_{ii}(x,x)/\mathrm{i} \,
      \bigr ] \>.
\end{split}
\end{equation}

The two-point vertex functions are defined by:
\begin{equation}\label{gold.e:twopointvertexdef}
   \Gamma_{ab}^{(2)}(x,x')
   \ = \
      - \frac{\delta^2 \Gamma[\phi]}
          {\delta \phi_a(x)\,\delta \phi_b(x')}
   \ = \
      \frac{\delta j_a(x)}{\delta \phi_b(x')} \>,
\end{equation}
so that by differentiating \eqref{gold.e:onepointvertexeq}, we find:
\begin{equation}\label{gold.e:twopointvertexeq}
   \Gamma_{ab}^{(2)}(x,x')
   \ = \
      G^{-1}_{0\,ab}(x,x')
   +
   \Sigma_{ab}(x,x') \>,
\end{equation}
where
\begin{align}
   G^{-1}_{0\,ab}(x,x')
   \ = \
   &   \bigl [ \,
      \Delta_{ab}^{-1}(x)
      +
      \gamma_{abc} \, \phi_c(x) \,
   \bigr ] \, \delta(x,x')
   \\
   \ = \
   &
   \begin{pmatrix}
      D^{-1}_0(x,x')   & \bar{K}_{0\,j}^{-1}(x,x') \\
      K_{0\,i}^{-1}(x,x') & G_{0\,i j}^{-1}(x,x')
   \end{pmatrix}  \>,
   \label{e:ginvdef}
   \notag
\end{align}
with
\begin{align*}
   D_0^{-1}(x,x')
   \ = \
   &
      - \, g \, \delta(x,x') \>,
   \\
   G_{0\,ij}^{-1}[\chi](x,x')
   \ = \
   &
      [ \, \Box + \chi(x) \, ] \, \delta_{ij} \delta(x,x') \>,
   \\
   K_{0\,i}^{-1}[\phi](x,x')
   \ = \
   &
      \bar{K}_{0\,i}^{-1}[\phi](x,x')
   \ = \
      \phi_i(x) \, \delta(x,x') \>.
\end{align*}
and
\begin{align}
   \Sigma_{ab}(x,x')
   \ = \
   &   \frac{1}{2i} \,
   \gamma_{abc} \,
   \frac{\delta G_{bc}(x,x)}{\delta \phi_b(x')} \>.
   \notag \\
   \ = \
   &
   \begin{pmatrix}
      \Pi(x,x')          & \Omega_j(x,x') \\
      \bar\Omega_i(x,x') & \Sigma_{ij}(x,x')
   \end{pmatrix}
   \>.
\end{align}
The two-point vertex and Green's functions are inverses of each
other:
\begin{equation}\label{gold.e:identity}
   \int \mathrm{d}^d x' \
   \Gamma_{ab}^{(2)}(x,x') \,
   G_{bc}(x',x'')
   \ = \
      \delta_{ac} \delta(x,x'') \>,
\end{equation}
from which we find schematically that
\begin{equation}
   \frac{\delta G_{ab}}{\delta \phi_c}
   \ = \
   -
   G_{a a_1} \, G_{b b_1} \,
   \Gamma_{a_1,b_1,c}^{(3)} \>.
\end{equation}
where
\begin{equation}\label{gold.e:threepointvertexdef}
   \Gamma_{abc}^{(3)}(x,x',x'')
   \ = \
      -
   \frac{\delta^3 \, \Gamma[\phi]}
        {\delta \phi_a(x)\,\delta \phi_b(x')\,\delta \phi_c(x'')} \>.
\end{equation}
is the three-point vertex function. So the self-energy
$\Sigma_{ab}(x,x')$ can be written as:
\begin{equation}\label{gold.e:Sigmadef}
   \Sigma_{ab}(x,x')
   \ = \
      \frac{\mathrm{i}}{2} \,
   \gamma_{a a_1 b_1}
   G_{a_1 a_2}
   G_{b_1 b_2}
   \Gamma_{a_2,b_2,b} \>.
\end{equation}

Differentiating Eq.~\eqref{gold.e:twopointvertexeq} again with
respect to $\phi_c(x'')$ gives an equation for the three-point
vertex function:
\begin{multline}\label{gold.e:GammaDGamma}
   \Gamma_{abc}^{(3)}(x,x',x'')
   \ = \
      \gamma_{abc} \, \delta(x,x') \, \delta(x,x'')
   \\
   +
   \Delta \Gamma_{abc}^{(3)}(x,x',x'') \>,
\end{multline}
where
\begin{equation}\label{gold.e:DGamma}
   \Delta \Gamma_{abc}^{(3)}(x,x',x'')
   \ = \
   \frac{ \delta \, \Sigma_{ab}(x,x') }
        { \delta \phi_{c}(x'') }
   \sim
   O(1/N) \>.
\end{equation}
For the purpose of renormalization it is useful to think of
Eq.~\eqref{gold.e:GammaDGamma} as of an identity
\begin{equation}
   \gamma
   \ = \
   \Gamma - \Delta \Gamma \equiv {\bar \Gamma}
   \>,
\end{equation}
since we have showed that both $\Gamma$ and $\Delta \Gamma$
renormalize the same way in our previous paper~\cite{renorm1}.

%
%

For the exact equations it is convenient to introduce the notations
\begin{align}
   \Gamma_{ab}^{(2)}(x,x')
   \ = \
      \begin{pmatrix}
      D_2^-1(x,x')         & \Xi_{j}(x,x') \\
      \bar{\Xi}_i(x,x') & G_{2,\, i j}^{-1}(x,x')
   \end{pmatrix}  \>,
\label{eq:ginvmat}
\end{align}
such that
\begin{subequations} 
\begin{align*}
   \Gamma^{(2)}_{00} & \equiv D_2^{-1}(x,x') = D_0^{-1}(x,x') + \Pi(x,x')\>, \\
   \Gamma^{(2)}_{ij} & \equiv G_{2,\, ij}^{-1}(x,x') = G_{0,\, ij}^{-1}(x,x') + \Sigma_{ij}(x,x') \>, \\
   \Gamma^{(2)}_{0j} & \equiv \Xi_{j}(x,x') = K_{0,\, j}^{-1}(x,x') + \Omega_j(x,x')
   \>, \\
   \Gamma^{(2)}_{i0} & \equiv \bar \Xi_{i}(x,x') = K_{0,\, i}^{-1}(x,x') + \bar \Omega_i(x,x')
   \>.
\end{align*}
\end{subequations}
In the homogeneous vacuum we can invert these equations in momentum
space to obtain schematically
\begin{align}
   D_2^{-1} D
    + \Xi_m \bar K_m
   & = \delta_{\C} \>,
   \notag \\
   \bar \Xi_i D + G_{2,\, im}^{-1} \bar K_m
   & = 0 \>,
   \\ \notag
   D_2^{-1} K_j + \Xi_m G_{mj}
   & = 0 \>,
   \\ \notag
   \bar \Xi_i K_j
    + G_{2,\, im}^{-1} G_{mj}
   & = \delta_{ij} \delta_{\C} \>.
   \notag
\end{align}
We find:
\begin{align}
   K_i
   = \bar K_i
   & = - D_2 \Xi_m G_{mi}
     = - G_{2,\, im} \bar \Xi_m D
   \>,
   \\
   D
   & = - g + g \, \bar \Pi \, D
   \>,
   \\
   G_{ij}
   & = G_0 \, \delta_{ij} - G_0 \bar \Sigma_{in} G_{nj}
   \>,
\end{align}
where we have introduced the notations
\begin{align}
   \bar \Sigma_{ij} =
   & \Sigma_{ij} - \bar \Xi_i D_2 \Xi_j
   \>,
   \\
   \bar \Pi =
   & \Pi - \Xi_m G_{2,\, mn} \bar \Xi_n
   \>.
\end{align}

The above equations are, in principle, exact. In practice, however,
the exact S-D hierarchy of equations needs to be truncated. Two
approximation schemes have been developed in the past few years: the
bare vertex approximation (BVA)~\cite{BVA}, where the resulting
dynamics is based on ignoring vertex corrections (i.e.~$\Gamma
\equiv \gamma$), and the 2PI--1/N expansion~\cite{2PI_1N}, where one
further ignores terms of order $1/N^2$.

%
%

In this paper, it is useful to define renormalization at $p^2=0$,
for the vacuum sector. As shown in our previous
paper~\cite{renorm1}, $\Sigma(p^2)$ is quadratically divergent and
requires two subtractions. Expanding about $p^2=0$, we have
\begin{equation}
   \Sigma(p^2) = \Sigma(0) + \Sigma_1 p^2 + \Sigma^{[\mathrm{sub} 2]}(p^2)
   \>,
\end{equation}
where $\Sigma_1 = \frac{\mathrm{d}\Sigma}{\mathrm{d}p^2} |_{p^2=0}$,
and $\Sigma^{[\mathrm{sub} 2]} \propto p^4$ as $p^2 \rightarrow 0$.
Then, the wave function renormalization constant is introduced as
\begin{equation}
   Z_2^{-1} = - \frac{dG^{-1}(p^2)}{dp^2} \biggr |_{p^2 = 0}
   \ = \ 1 - \Sigma_1
   \>.
\end{equation}
The vacuum renormalized mass parameter is defined as
\begin{equation}
   M^2 \ = \ Z_2 \ [ \chi + \Sigma(0) ]
   \>.
\end{equation}

The vertex renormalization constant $Z_1$ is equal to $Z_2$ by a
Ward-like identity and is defined by
\begin{equation}
   Z_1^{-1} =
   \Gamma(p,p) |_{p^2=0}
   \ = \
   1 + \frac{\partial \Sigma(p^2)}{\partial \chi} \biggl | _{p^2=0}
   \>,
\end{equation}
and $\Gamma_R(p,q) = Z_1 \Gamma(p,q)$. We have shown in the vacuum
sector, that
\begin{align}
   G_R^{-1}(p^2)
   & \ = \ Z_2 \ G^{-1}(p^2)
   \\ \notag &
   \ = \
   p^2 + M^2 + \Sigma_R^{[\mathrm{sub} 2]}(p^2)
   \>,
\end{align}
where $\Sigma_R^{[\mathrm{sub} 2]}(p^2)$ is explicitly finite and
written only in terms of renormalized Green's functions and
renormalized vertex functions.

Also, in 3+1 dimensions, coupling constant renormalization is
needed. Since the renormalized coupling constant $g_R$ is the
negative of the inverse $\chi$ propagator at $p^2=0$, and is a
renormalization group invariant, one can obtain a finite equation
for $D^{-1}$ with the following single subtraction
\begin{equation}
   D^{-1}(p^2) = -\frac{1}{g_r} + \Pi^{[\mathrm{sub} 1]} (p^2)
   \>,
\end{equation}
with
\begin{equation}
   \Pi^{[\mathrm{sub} 1]} (p^2) \ = \
   \Pi(p^2) \ - \ \Pi(0)
   \>.
\end{equation}
What we showed in our previous paper~\cite{renorm1} is that one can
write:
\begin{equation}
   \Gamma_R(p,q) = 1 + \Delta \Gamma_R^{[sub1]} (p,q)
   \>,
\end{equation}
where the second term is finite, renormalized, and of order $1/N$.

%
%
\section{2PI--1/N expansion}

\begin{figure}[b]
   \includegraphics[width=2.in]{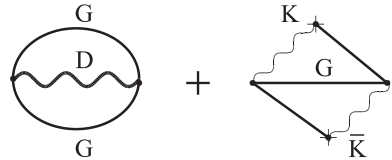}
   \caption{\label{f:fig1}
   Graphs included in the 2PI effective action
   $\Gamma_2[G]$.}
\end{figure}

Next we want to compare these exact results with the S-D equations
coming from the 2PI--1/N approximation.  Now we have that the
generating functional is given by:
\begin{align}
   \Gamma[\phi_a,G]
   \ = \ &
   S_{\text{cl}}[\phi_a]
   +
   \frac{\mathrm{i}}{2} {\rm Tr} \ln [ \, G^{-1} \, ]
   \notag \\ &
   +
   \frac{\mathrm{i}}{2} {\rm Tr} [ \, G_0^{-1} \, G \, ]
   +
   \Gamma_2[G] \>,
   \label{e:BVAaction}
\end{align}
where $ \Gamma_2[G]$ is the generating functional of the 2PI
graphs~\cite{2PI_CJT}, and $S_{\text{cl}}[\phi_a]$ is the classical
action in Minkowski space. The approximations we are studying
include only the two-loop contributions to $\Gamma_{2}$ (see
Fig~\ref{f:fig1}).

The exact equations following from the effective action
Eq.~\eqref{e:BVAaction}, are the same as Eqs.~\eqref{eq_motion} and
\eqref{gold.e:twopointvertexeq} listed above, with the Green's
function $G^{-1}_{0\,a b}[\phi](x,x')$ defined as
\begin{equation}
   G_{0\,a b}^{-1}[\phi](x,x')
   \ = \
   - \frac{\delta^2 S_{\text{cl}} }
          { \delta \phi_{a}(x) \ \delta \phi_{b}(x') }
   \>,
\end{equation}
and the self-energy
\begin{equation}
   \Sigma_{a b}(x,x')
   \ = \
   \frac{2}{\mathrm{i}} \
   \frac{\delta \Gamma_2[G]}{\delta G_{ab}(x,x')}
   \>.
   \label{e:Sigmasdefs}
\end{equation}

%
%

In the 2PI--1/N we keep in $\Gamma_2[G]$ only the first of the
two graphs shown in Fig.~\ref{f:fig1}, which is explicitly
\begin{align}
   \label{e:Gamma2}
   \Gamma_2[G]
   \ = \
      - \frac{1}{4}
   \iint \mathrm{d}^d x \ \mathrm{d}^d y \
      G_{ij}(x,y) G_{ji}(y,x) D(x,y)
   \>.
\end{align}
The self-energy, given in Eq.~\eqref{e:Sigmasdefs}, then reduces
to:
\begin{align}
   \Pi(x,x') \ = \
   & \frac{\mathrm{i}}{2} \
      G_{mn}(x,x') \, G_{mn}(x,x')
   \>,
   \notag \\
   \Omega_i(x,x') \ = \
   & 0
       \>,
   \label{e:SigmasBVA} \\
   \bar\Omega_i(x,x') \ = \
   & 0
   \>,
   \notag \\
   \Sigma_{ij}(x,x') \ = \
   & \mathrm{i} \
         G_{i j}(x,x')\, D(x,x')
   \>.
   \notag
\end{align}

In the homogeneous case we will use the O(N) symmetry to choose the
symmetry breaking direction to be in the direction N. In that case
only $\langle \phi_N \rangle \neq 0$. This means that $G_{ij}$ is
diagonal in general, and only the fields $\chi$ and $\phi_N \equiv
\sigma$ mix. To determine the $\sigma$ mass one has to just
diagonalize a $2 \times 2$ matrix. There is no mixing between $\chi$
and the $\phi_i$ where $i  < N$. Thus $K_N$ is the only non zero
entry to the mixed propagator $K_i$ and $ \Xi_{j}(x,x')= \delta_{jN}
\phi_N \delta(x,x')$. Let us look at the momentum space equations.
The integral equations for $D$ and $G_{ij}$ are now
\begin{align}
   D
   & = - g + g \, \bar \Pi \, D
   \>,
   \\
   G_{ij}
   & = G_0 \, \delta_{ij} - G_0 \bar \Sigma_{in} G_{nj}
   \>,
\end{align}
with
\begin{align}
   \bar \Sigma_{ij} =
   &  \Sigma_{ij} - \delta_{iN} \phi_N \bar   D_2 \delta_{jN} \phi_N
   \>,
   \\
   \bar \Pi =
   &  \Pi -  \delta_{mN} \phi_N G_{2,\, mn}  \delta_{nN} \phi_N
   \>.
\end{align}
Iterating the equation for $G$ shows it is diagonal and only
$G_{NN}$ is different from $G_{2\,NN}$. For the self-energy only the
$NN$ component is modified from the unbroken case. We also have that
\begin{equation}
   D^{-1} = D_2^{-1}- \phi_N G_{2,NN} \phi_N
   \>.
\end{equation}
This difference will be important when we discuss Goldstone's
theorem. It follows immediately that, in momentum space, we have
\begin{align}
   D^{-1}(p^2)
   \ = \
   &    - \frac{1}{g} + \bar \Pi(p^2)
   \>,
   \\
   G_{ij}^{-1}(p^2)
   \ = \
   &    ( -p^2 + \chi ) \, \delta_{ij} + \bar \Sigma_{ij}(p^2)
   \>,
\end{align}
and
\begin{align}
   K_i(p^2) = \bar K_i(p^2)
   & = - \delta_{iN}\phi_N  D_2 G_{NN}
   \>.
\end{align}
Now the correction to $\Pi(p^2)$ goes like $1/p^2$ so this is
irrelevant at high momentum. The correction to $\Sigma(p^2)$ goes
like $1/\ln p^2$ so this is also negligible compared to $\ln p^2$.
This implies that the renormalizability is not changed by symmetry
breaking. Also, since $G$ and $\phi^2$ renormalize the same way the
multiplicative renormalization does not change. Introducing the
notations $\phi_N = \sigma$ and $\phi_{i\, (i \neq N)} = \pi_i$ and
letting $G_{ij} = G \ \delta_{ij}$ we have that the inverse
propagator for the  $\pi$ mesons is
\begin{equation}
   G^{-1}_{\pi \pi}(p^2) \ = \
   -p^2 + \chi + \Sigma(p^2)
   \>,
\end{equation}
and for the $\sigma$ meson we have instead
\begin{equation}
   G^{-1}_{\sigma \sigma}(p^2) \ = \
   -p^2 + \chi + \Sigma(p^2) - |\phi|^2 D_2(p^2)
   \>.
\end{equation}

%
%
\section{Goldstone Theorem}

The one-point function equation in an external source is
\begin{equation}
   \bigl [ \Box+ \chi(x) \bigr ] \, \phi_i(x) + K_i(x,x)/i = j_i(x)
   \>.
\end{equation}
This is to be interpreted as
\begin{equation}
   G_{3\,ij}^{-1} (x,x') \ \phi_j(x')
   \ = \
   j_i(x)
   \>.
\end{equation}
where
\begin{equation}
   G_{3\,ij}^{-1}(x,x')
   =
   \bigl [ \Box + \chi(x) \bigr ] \delta_{ij} \delta (x-x') + \Sigma_{3\,ij}(x,x')
   \>.
\end{equation}
Now since
\begin{equation}
   K_i (x,x')
   \ = \
   - \, D_2(x,x') \phi_j (x') G_{ji}(x,x')
   \>,
\end{equation}
we see that
\begin{equation}
   \Sigma_{3\,ij}(x,x')
   \ = \
   \mathrm{i} \ G_{ij}(x,x') D_2(x,x')
   \>.
\end{equation}
Thus apart from $D \rightarrow D_2$, this is exactly the self-energy
$\Sigma_{ij}(x,x')$~! Thus $G_3$ is made finite by exactly the same
2 subtractions of wave function renormalization and mass
renormalization as the full $G$. The renormalized one-point function
equation is then
\begin{equation}
   G_{3 R}^{-1} (x,x') \ \phi_R(x')
   \ = \
   0
   \>.
\end{equation}
In momentum space, in the vacuum, we have
\begin{equation}
   G_{3 R}^{-1}(p^2) \ = \
   -p^2 + M^2_3 (0) + \Sigma_{3R}^{\mathrm{sub\,2}}(p^2)
   \>,
\end{equation}
where the self-energy is subtracted twice at $p^2=0$. We also have
\begin{equation}
   M^2_3
   \ = \
   Z_2 \ \bigl [ \chi + \Sigma_3 (0) \bigr ]
   \>.
\end{equation}

The condition for broken symmetry is that
\begin{equation}
   \bigl ( \chi \delta_{ij} + \Sigma_{3\,ij} \bigr ) \ \phi_i \ = \ 0
   \>.
\end{equation}
Choosing  the direction of the expectation value $\langle{\vec \phi}
\rangle$ to define the  $i=N$ direction we have
\begin{equation}
   \chi + \Sigma_{3\,NN} \ = \ 0
   \>,
\label{4.10}
\end{equation}
for spontaneous symmetry breakdown. We need to ask whether this
insures N-1 Goldstone bosons (see also previous discussions on this
topic in~\cite{baym_77,Aarts_66,hees_02a,hees_02b}).

Now the N-1 would be Goldstone bosons come from the inverse
propagator $G^{-1}_{\pi \pi}$, which after renormalization at
$p^2=0$ gives
\begin{equation}
   G_R^{-1}(p^2) \ = \ - p^2 + M^2(0) + \Sigma_R^{\mathrm{sub\,2}}(p^2)
   \>,
\end{equation}
with $\Sigma^{sub2}_R$ (the twice subtracted at $p^2=0$ renormalized
self-energy) proportional to $ p^4$ at small $p^2$. The condition
for a Goldstone theorem is that $G_R^{-1} = a p^2$ for small $p^2$
so that there is a zero mass pole in the propagator. This requires
\begin{equation}
   M^2(0) \ = \ Z_2 \ \bigl [ \chi + \Sigma(0) \bigr ] \ = \ 0
   \>.
\end{equation}
However the condition for broken symmetry is that
\begin{equation}
   M^2_3 \ = \ Z_2 \ \bigl [ \chi + \Sigma_3 (0) \bigr ] \ = \ 0
   \>.
\end{equation}
The difference between $\Sigma$ and $\Sigma_3$ is of order 1/N, and
is proportional to $\langle \phi \rangle^2$.

If we want to preserve the Goldstone theorem in our dynamical
simulations we could use $D_2$ and not the full $D$ in our update
equations for the self-energy.  However this would then violate
energy conservation (by terms of order 1/N) previously guaranteed by
the use of the effective action. Note, however, that if were only
interested in O(4) symmetric initial condition, but having Goldstone
particles, the strategy of Arizabalaga \emph{et al} works perfectly.
By first choosing $\langle \phi \rangle$ small, but not zero,
Eq.~\eqref{4.10} must be satisfied. Taking the limit $\langle \phi
\rangle$ goes to zero later, the difference between $\Sigma$ and
$\Sigma_3$ vanishes, and we have no conceptual problem. The
difficulty only arises when one is interested in non O(N) symmetric
initial conditions for the expectation value of $\phi$.

In leading order 1/N, the self-energy $\Sigma$ is zero and the
condition for symmetry breakdown is then  $\chi =0$ which
automatically leads to N-1 Goldstone particles, and there is no
problem with the Goldstone theorem. (This fact has been verified to
order 1/N$^2$ in a direct 1/N expansion by Binoth \emph{et
al}~\cite{Binoth}.) As for the mass of the $\sigma$ meson one has
that
\begin{equation}
   m_\sigma^2(0) - m_\pi^2(0)
   \ = \
   - Z_2 \ \phi^2 \ D_2(0) \ \equiv \ g_R \phi_R^2/2
   \>.
\end{equation}
This is the renormalized version of what happens in the classical
theory. To make a realistic model of pions, one has to {
\em{explicitly} }break the O(4) symmetry by setting $j_0 = H $ as in
Ref.~\cite{DCC}. Doing this the quantum field equation for the
$\sigma$ field becomes
\begin{equation}
   \bigl [ \Box + \chi \bigr ] \, \sigma \ = \ H
   \>.
\end{equation}
Therefore, $H$ is renormalized the same way as $\sigma$, and the
renormalized PCAC equation coming from
\begin{equation}
   A_\mu^i \ = \ \pi^i \partial_\mu \sigma - \sigma \partial_\mu \pi^i
   \>, \quad  (i = 1 \ldots N-1)
   \>,
\end{equation}
becomes
\begin{equation}
   \partial^\mu A^i_{R~\mu} (x) \ = \ H_R \pi^i_R(x)
   \>,
\end{equation}
with $H_R = f_\pi m_\pi^2$. As long as the pion mass generated from
the breakdown of the Goldstone theorem is small compared to the mass
coming from the explicit symmetry breakdown, the violation of the
Goldstone theorem by this approximation will not be important in
dynamical simulations of an effective theory of disoriented chiral
condensates.

One way to ``solve'' the Goldstone problem is to introduce an
``improved'' action~\cite{hees_02a,Aarts_66}
\begin{equation}
   \Gamma^\ast [\phi] \ = \ \Gamma[\phi, \chi[\phi] , G[\phi]]
   \>.
\end{equation}
The second derivative of this action is guaranteed to satisfy the
Goldstone theorem by construction. Because of the O(N) symmetry
$\Gamma^\ast$ is only a function of $\phi \cdot \phi \equiv \phi^2$.
Thus the condition for a minimum is
\begin{equation}
   \frac {\partial \Gamma^\ast}{\partial \phi_i}
   \ = \ 2 \ \frac{\partial \Gamma^\ast}{\partial \phi^2} \ \phi_i =0
   \>.
\end{equation}
So that for $\langle \phi_i \rangle \neq 0$, we have
\begin{equation}
   \Gamma^{\ast \prime} \ = \ \frac {\partial \Gamma^{\ast}}{\partial \phi^2}=0
   \>,
\end{equation}
at the minimum. The inverse
propagator is now
\begin{align}
   G^{-1}_{ij}
   & = \frac {\partial^2 \Gamma^{ \ast}}{\partial \phi_i \partial \phi_j }
   \\ \notag
   & = 2 \ \Gamma^{\ast \prime} \ [ \delta_{ij} - \phi_i \phi_j/\phi^2]
     +  (2 \Gamma^{ \ast \prime } + 4 \Gamma^{\ast \prime \prime }) \phi_i \phi_j/\phi^2
   \>.
\end{align}
From this equation one infers that the transverse degrees of freedom
are massless and the longitudinal ones are not.  The construction of
$\Gamma^\ast$ though feasible in $3+1$ dimensions in static cases,
is not at present numerically feasible in the dynamical case where
one has to solve further Bethe-Salpeter equations for the
three-point vertex functions. The details of the construction of
$\Gamma^\ast$ are found in~\cite{hees_02a}.

Before closing, let us remark that in a recent paper, Ivanov
\emph{et al}~\cite{knoll05} have proposed a new way of circumventing
the violation of Goldstone's theorem, at leading order, in the
simpler Hartree approximation. Specifically, these authors have
outlined a modified self-consistent Hartree approximation, which
preserves features present in the $\phi$-derivable approach, such as
energy conservation and thermodynamic consistency. By adding terms
to the 2PI generating functional which vanish when the symmetry is
restored, their approach explicitly enforces the Goldstone theorem.
This may be a promising approach, and it will be interesting to see
if the same strategy can be pursued at next-to-leading order in 1/N.

%
%
\section{Conclusions}

In what is a follow-up to our previous paper~\cite{renorm1}, in
which we have discussed the renormalization of the symmetric O(N)
model, $\phi = 0$, to next-to-leading order in 1/N, in the S-D
framework, in this paper we have shown that the 2PI--1/N expansion
of the O(N) model in the homogeneous broken symmetry vacuum is also
renormalizable to order 1/N. We have derived finite equations for
the renormalized Green's functions, and shown that Goldstone's
theorem is violated. We have briefly discussed some current ideas
about how to circumvent this problem. Our major interest here was to
obtain finite renormalized equations for numerical simulations of
O(4) model dynamics. To make a realistic model of the time evolution
of the chiral phase transition with physical $\pi$ mesons requires
introduction of explicit symmetry breakdown~\cite{DCC}, which will
make the violation of the Goldstone theorem unimportant in
phenomenological applications, when compared with the mass generated
by the explicit breaking of the O(4) symmetry. The renormalization
presented here is easily generalized to the time-dependent equations
and we are in the process of reinvestigating the problem of
disoriented chiral condensates using the O(4) model in the 2PI--1/N
expansion.

%
%

\newpage

\begin{acknowledgments}
We would like to thank the Santa Fe Institute for its
hospitality during the completion of this work. We would also like
to thank the DOE and the NSF for their partial support of this work.
\end{acknowledgments}

%
%

%
%

\end{document}